# Axisymmetric model for 1-color laser filament THz emission

Sean D. McGuire[1,*] and Mikhail N. Shneider[2]

[1]Université Paris-Saclay, CNRS, CentraleSupélec, Laboratoire EM2C, 91190, Gif-sur-Yvette, France.
[2]Princeton University, Princeton, New Jersey 08544, USA

**ABSTRACT**. A 3D axisymmetric model for laser beam propagation in air is proposed for use with THz emission calculations. The model accounts for nonlinear propagation of the femtosecond pulse that generates the filament (Kerr self-focusing and plasma defocusing). The formalism proposed can be applied across a broad range of experimental parameters – focal lengths, beam profiles, and intensities. We adapt existing models of 1-color THz emission for use with the beam propagation model to calculate THz emission signatures. Two of these THz emission models are used - a 1D model and a more detailed 3D axisymmetric model. For the test case considered, nonlinear propagation dynamics have a large impact on the amplitude of the THz signal, but only a minimal impact on the frequency content and angular distribution of the signal.

## I. INTRODUCTION.

Femtosecond filaments occur when a high intensity ($> 10^{13}\ W/cm^2$) short pulse laser ($< 100\ fs$) generates a weakly ionized plasma via photoionization [1, 2]. Couairon and Mysyrowicz provide a detailed overview of femtosecond pulse propagation in gases [2]. Defocusing of the optical beam occurs due to the filament ionization. This effect counters the tendency of the optical pulse to self-focus due to the nonlinear Kerr effect. The interplay between these two nonlinear propagation effects limits the intensity of the focused beam and the level of photoionization. As a result, electron densities are limited to $n_e < 10^{18} cm^{-3}$ in air. An interesting phenomenon associated with femtosecond filaments is their emission of radiation in the GHz and THz frequency ranges (FIG 1) [3-5]. Emission in the THz frequency range was first observed by Hamster in the early 1990's [4, 6, 7] and generated significant interest due to the notable lack of practical emission sources in this frequency range – the so-called *THz gap* [8]. The angular distribution of the THz emission is case dependent but generally occurs either in the forward or side directions [5, 7]. Recent work has focused on the use of a 2-color optical field for generating THz emission as this produces a much larger THz signal compared to the 1-color mechanism studied by Hamster et al – Tailliez et al [9] report conversion efficiencies for THz emission on the order of $10^{-2}$ for the 2-color approach, compared to efficiencies $< 10^{-8}$ for the 1-color approach reported by Sprangle et al [10]. Additional approaches involving optical mixing have also been proposed [11].

Nonlinear laser beam propagation effects such as Kerr self-focusing and plasma defocusing can significantly alter the intensity distribution in the focal zone. Consequently, such effects should be accounted for when using high laser intensities ($\gtrsim 1.0 \times 10^{13}\ W/cm^2$ in air). Various methods exist for modeling nonlinear propagation effects. Sprangle et al perform numerical simulations of THz emission from a femtosecond filament using a 1-color optical pulse [10], and account for nonlinear propagation affects using a *Source Dependent Expansion* method for modeling the evolution of the laser pulse. This method assumes a Gaussian transverse beam profile. Prost *et al* [12] and Babushkin *et al* [13] use a unidirectional pulse propagation equation in their work. In this paper, we use a 3D axisymmetric beam propagation model based upon the use of Hankel transforms and apply it to calculations of THz emission via the 1-color *Transition-Cherenkov* THz emission mechanism identified in the literature [5, 14, 15].

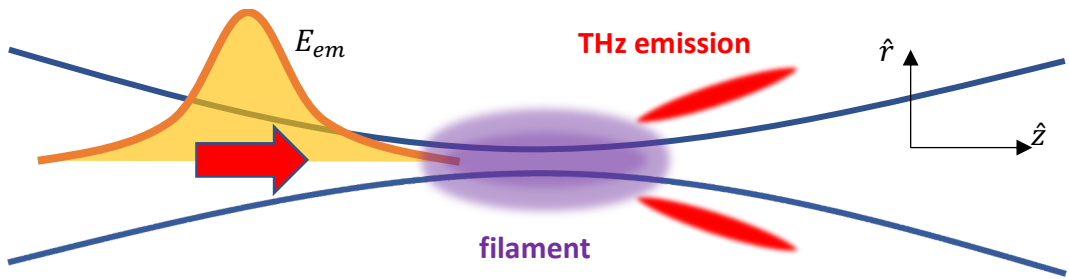

FIG 1: Experimental configuration used for the generation of THz radiation from a single filament. The direction of the emitted THz radiation depends upon the experimental parameters used but generally falls between the forward ($+\hat{z}$) and side directions.

*Contact author: sean.mc-guire@centralesupelec.fr

The proposed beam propagation model is coupled to an existing model of the 1-color Transition-Cherenkov THz emission available in the literature, which takes the laser electric field strength as an input and provides the electron current generated as an output. Here, we consider the case of a filament generated in ambient air. The resulting THz emission is then calculated using a classical solution to Maxwell's equations. The impact of nonlinear propagation effects on the THz signal amplitude, frequency content, and angular emission profile is evaluated. For the case considered here, nonlinear propagation effects have a large impact on the amplitude of the THz signal but only a minor effect on the frequency content and angular distribution of the emission. Finally, a simplified 1D model is proposed which, when used in conjunction with the beam propagation model, reproduces reasonably well all characteristics of the more detailed calculations (amplitude, frequency content, and angular emission profile).

Section II of this paper introduces the beam propagation model used for all calculations. Section III introduces two models for the Transition-Cherenkov emission mechanism – a 3D model axisymmetric and 1D model – which are both coupled to the beam propagation model. Section IV shows the impact of nonlinear propagation effects on the THz signal and provides a comparison between the predictions of the two THz emission models.

## II. BEAM PROPAGATION MODEL

In a cylindrical coordinate system, the spatial and temporal evolution of the propagating electric field can be expressed as follows:

$$E(r,z,t) = Re[\bar{E}(r,t-z/c)e^{i(k_L z - \omega_L t)}] \quad 1$$

where $k_L$ and $\omega_L$ are the laser wavenumber and angular frequency, respectively, and $\omega_L/k_L = c$. The beam propagation model assumes that the slowly varying envelope approximation applies. Thus, time derivatives of $\bar{E}(r, t-z/c)$ are neglected and the evolution of the total electric field $E(r,z,t)$ is governed by the homogeneous Helmholtz equation.

$$\nabla^2 E(r,z,t) + \frac{\omega_L^2 n^2}{c^2} E(r,z,t) = 0 \quad 2$$

where $n$ is the index of refraction. Eqn. 2 neglects terms involving spatial gradients and time derivatives of the index of refraction, which is valid here given that these modulations occur over distances much larger than the laser wavelength and time periods much longer than the optical period [16]. The slowly varying envelope approximation also implies that spatial derivatives of $\bar{E}(r, t-z/c)$ along the axial coordinate $z$ can be neglected upon substitution of Eqn. 1 into Eqn. 2. Finally, Eqn. 2 is solved only at the laser frequency $\omega_L$, meaning that dispersion in the optical pulse is neglected.

The index of refraction is modulated by the Kerr effect ($\Delta n_{kerr} = n'I$, where $I$ is the intensity and $n' = 5.0 \times 10^{-23} m^2/W$ in air [17]) and electron charge density ($\Delta n_{pl} = \sqrt{1 - \omega_p^2/\omega^2} - 1$). The operator splitting method of Feit and Fleck is used to account for these variations and numerically solve Eqn. 2 [18]. This approach has been previously used to describe pulse propagation through plasmas by New-Tolley, Shneider and Miles [19]. The method consists in the iterative application of a two-step process. First, the field is propagated forward a short distance. Second, a spatially dependent phase factor is then applied to account for variations in the index of refraction.

For the propagation step, the formalism developed by Guizar-Sicairos and Gutiérrez-Vega is used [20]. This approach is based around the use of the Hankel transform to model the axi-symmetric beam propagation. Given a transverse electric field profile $E(r)$, the forward and inverse *p*-th order Hankel transforms read as follows:

$$\hat{E}(\nu) = 2\pi \int_0^\infty E(r) J_p(2\pi\nu r) \, r dr \quad 3$$

$$E(r) = 2\pi \int_0^\infty \hat{E}(\nu) J_p(2\pi\nu r) \, \nu d\nu \quad 4$$

where $J_p$ is the *p*-th order Bessel function and $\nu$ is the spatial frequency. One advantage of this approach is that it does not impose any assumptions on the transverse beam profile. Guizar-Sicairos and Gutiérrez-Vega, for example, use this algorithm to simulate the focusing of a Bessel beam [20].

The above considerations lead to the following sequence of operations, which are repeated iteratively.
1. The forward Hankel transform is applied to the radial profile of the electric field and provides the transformed electric field as a function of radial spatial frequency.
2. A multiplicative propagation factor is applied in the spatial frequency domain.

3. The inverse Hankel transform is applied to transform the spatial frequency domain back to the radial coordinate domain.
4. A multiplicative phase factor accounting for index of refraction variations due to Kerr self-focusing and plasma defocusing is applied.

This sequence of steps is applied at every axial position. Correctly capturing all spatial frequencies requires a highly resolved radial grid. These considerations can result in lengthy computations when many mesh points are required along the radial axis – namely, when tightly focusing the laser beam. For the case considered in this paper, the beam propagation calculations were implemented in python and took several hours.

Note that the model is 3D but uses a cylindrical coordinate system and assumes axisymmetry. We do not account for the reduction in the pulse energy due to photoionization processes. For the cases considered in this paper, the total energy absorbed amounts to less than 1% of the total pulse energy. However, such absorption could be accounted for within the context of the operator splitting method. Finally, while a Gaussian radial beam profile is assumed, this is not a necessary assumption.

### III. CURRENT DYNAMICS AND EMISSION MODELS

To simulate THz emission from a femtosecond filament, the beam propagation model must be coupled to a separate model which describes the current dynamics and the resulting emission. In this paper, two models are considered and introduced in the following subsections. The first is a fully *3D axisymmetric* model. The second uses the 3D axisymmetric beam propagation model presented above, but simplified 1D expressions for the current and emission models. This second model will be referred to as the *quasi-1D* model. For both models, a linearly polarized optical pulse is assumed.

#### IIIA. 3D Axisymmetric THz emission model

The current dynamics are described using the model of Thiele *et al* [5], who give a detailed derivation starting from kinetic theory and validate the model against PIC simulations. The model uses a multiple scale expansion wherein the electron density, current density, and electromagnetic fields are expanded as part of a perturbation series. The multiple scale expansion is shown to be valid for nonrelativistic laser pulses and results in separate equations for each order of the perturbation expansion. The equations of relevance for this work, labeled according to their order ($\epsilon^0, \epsilon^1, \epsilon^2, ...$), are as follows:

$$\epsilon^0 \qquad \partial_t n_0 = S \qquad \qquad 5$$

$$\epsilon^1 \qquad \partial_t \boldsymbol{J_1} + \nu_e \boldsymbol{J_1} = \frac{e^2}{m_e} n_0 \boldsymbol{E_1} \qquad 6$$

$$\epsilon^2 \qquad \partial_t \boldsymbol{J_2} + \nu_e \boldsymbol{J_2} = \frac{e^2}{m_e} n_0 \boldsymbol{E_2} + \boldsymbol{l_2} \qquad 7$$

where

$$\boldsymbol{l_2} = \frac{n_0}{2e} \boldsymbol{\nabla} \left|\frac{\boldsymbol{J_1}}{n_0}\right|^2 + \\ \frac{\boldsymbol{J_1}}{e} \times \boldsymbol{\nabla} \times \int_{-\infty}^{t} \frac{\boldsymbol{J_1}}{n_0}\left(\nu_e + \frac{\partial_{t'} n_0}{n_0}\right) dt' + \\ \frac{(\nu_e + \partial_t)}{e n_0}\left(\boldsymbol{J_1} \int_{-\infty}^{t} \boldsymbol{\nabla} \cdot \boldsymbol{J_1}\, dt'\right) + \\ \frac{2e}{3 m_e} \boldsymbol{\nabla}(n_0 \mathcal{E}_{th}) \qquad 8$$

$\nu_e$ and $\mathcal{E}_{th}$ are the electron collision frequency and electron thermal energy, respectively. The bold variables are vectors in the above equations, and this is the convention that will be used in this paper. $\boldsymbol{J_1}$ and $\boldsymbol{J_2}$ are contributions to the total current (both contain displacement + conduction components) at first and second order, respectively. In other words, the total current $\boldsymbol{J} = \boldsymbol{J_1} + \boldsymbol{J_2} + \cdots$. Similarly, the electron density is $n = n_0 + n_1 + n_2 + \cdots$. Because the ions are treated as stationary, the ion density is equal to $n_0$. Thiele *et al* give additional expressions for calculating $\nu_e$ and $\mathcal{E}_{th}$. It is worth noting that Thiele *et al* take $\nu_e$ to be the electron-ion collision frequency, whereas here the total electron collision frequency is used (electron-ion plus electron-neutral). This is done because filaments in air remain weakly ionized. The electron-ion and electron-neutral collision frequencies are assumed to be constant for simplicity. These are calculated as described in Ref. [21] and fixed to $2.0 \times 10^{12}\ s^{-1}$ and $1 \times 10^{12}\ s^{-1}$, respectively, giving $\nu_e = 3.0 \times 10^{12}\ s^{-1}$. In general, the electric fields in the above equations are solved using Maxwell's equations and begin at first order ($\epsilon^1$). The electric field $\boldsymbol{E_1}$ includes the laser electric field, as well as any electric fields produced via $\boldsymbol{J_1}$. The *ionization current* mechanism of THz emission is accounted for in the dynamics of $\boldsymbol{J_1}$ as described by Eqn. 6. The *transition-Cherenkov* mechanism that is the focus of this paper is accounted for in the dynamics of $\boldsymbol{J_2}$ according to Eqn. 7. Prior work showed that the *transition-Cherenkov* mechanism is the dominant emission mechanism for pulse lengths above $25\ fs$ [5]. Therefore, the laser pulse length used in this work was chosen to be $50\ fs$.

The adaptation of the model proposed here uses a cylindrical coordinate system. Axisymmetry is assumed for the magnitude of the vectors $\boldsymbol{J_1}$ and $\boldsymbol{E_1}$, but not their direction which is fixed along the fixed axis of laser polarization. Axisymmetry is assumed for all other variables in Eqns. 5, 7, and 8. Certain current source terms on the right-side of Eqn. 8 are not compatible with the assumption of axisymmetry given that the directions of the vectors $J_1$ and $E_1$ are not axisymmetric. The third term on the right-side is not compatible – as a vector, this term is directed along $J_1$ which is directed along the laser polarization axis $E_1/\|E_1\|$ because of Eqn. 6. This term is therefore ignored. Meanwhile, as detailed in Appendix B, the second term on the right-side is only partially compatible – only the portion which is compatible is retained for the calculations in this paper. There are a few additional approximations used, and these are as follows:

- $E_1$ is taken to be the laser electric field – the electric field due to $J_1$ is neglected and the plasma is considered transparent to the high frequency optical field. This simplification has been made by other authors (e.g. Ref. [14]) and is discussed in Section II of Thiele et al [5]. Axisymmetric models – such as are being used here – cannot capture the contribution to $E_1$ from $J_1$ as the direction of $J_1$ is not axisymmetric. Therefore, though this hypothesis is supported by results in the literature, its validity cannot be tested here.
- The slowly varying envelop approximation is applied to $E_1$ and $J_1$.
- $E_2$ is determined using Poisson's equation : $\nabla \cdot E_2 = -(e/\varepsilon_0)(n_2 - n_0)$, where $n_2(r,t)$ is calculated directly from $J_2$ using the continuity equation. The full set of Maxwell's equations is not used. For the calculations considered in this paper, the force term $l_2$ on the right-hand side of Eqn. 7 is much larger in magnitude than the force term $(e^2 n_0 E_2/m_e)$ as the laser transits the domain. Thiele et al completely neglect $E_2$ in their calculation of THz spectra and, for the cases considered in their paper, find good agreement with PIC simulations [5].

The above approximations are applied to Eqns. 5 – 8 in Appendices A and B. The simplified equations are then numerically resolved to determine the current $J_2$. The rate of photoionization $S$ in Eqn. 5 is calculated using the ionization rates of Mishima et al for molecular nitrogen and oxygen at $800\ nm$ [22]. The air is taken to be composed of 80% $N_2$ and 20% $O_2$ by volume. The evolution in the electron temperature $\mathcal{E}_{th}$ is not shown in the figures below but increases to a maximum of about $3\ eV$.

Given the current dynamics calculated using the Thiele et al model with the above approximations, the emission from the $J_2$ current can be calculated. Eqn. 9.3 of Ref. [23] is used, which provides one Fourier component of the vector potential:

$$A(r,\omega) = \frac{\mu_0}{4\pi} \int J(r',\omega) \frac{e^{ik|r-r'|}}{|r-r'|} d^3 r' \quad 9$$

where $J(r,\omega)$ is the Fourier component of the current density and $k = \omega/c$. The electric field outside of the current source may then be calculated as in Ref. [23].

$$E(r,\omega) = i\frac{c}{k} \nabla \times \nabla \times A(r,\omega) \quad 10$$

The temporal dependence of the electric field may be recovered by performing an inverse Fourier transform of $E(r,\omega)$. Note that this emission calculation neglects the effects of dispersion and absorption of the THz emission signal within the filament, which can become important at frequencies lower than the plasma frequency depending upon the spatial extent of the filament [5].

### IIIB. Quasi-1D THz emission model

The quasi-1D model for current dynamics uses all the approximations invoked for the 3D axisymmetric model of current dynamics but adapts the equations to a 1D geometry, wherein only variations along the axis of laser propagation ($\hat{z}$) are accounted for. Appendix C derives a simple expression for the term $l_2$ in Eqn. 8. This expression is as follows:

$$l_2(\eta) = l_2(\eta)\,\hat{z} = -\frac{1}{e\varepsilon_0 c^2}\left(\frac{e^2}{m_e}\right)^2 \frac{1}{\omega_L^2 + \nu_e^2} \times$$
$$\left[\frac{n_0}{2}\frac{dI}{d\eta} + \nu_e n_0 I + I\frac{dn_0}{d\eta}\right]\hat{z} \quad 11$$

where $\eta = t - z/c$ and $I(\eta) = (1/2)c\epsilon_0 \bar{E}_1 \bar{E}_1^*$ is the intensity of the laser. $\omega_L$ is the laser carrier frequency. Here, the intensity at the focal point is used. The three terms in brackets account for the ponderomotive source, radiation pressure source, and ionization source, respectively. Eqn. 11 can be substituted into Eqn. 7 to provide an equation for the current $J_2$. The electric force term is ignored, as its inclusion incorrectly attributes radiation to a non-radiating current [5]. Eqn. 7 becomes $(\partial_t J_2 + \nu_e J_2 = l_2) * \hat{z}$. Finally, applying a Fourier transform gives:

$$\hat{J}_2(\omega) = \frac{\hat{l}_2(\omega)}{i\omega + \nu_e} \quad 12$$

where $\hat{J}_2(\omega)$ and $\hat{l}_2(\omega)$ are the Fourier transforms of $J_2(\eta)$ and $l_2(\eta)$, respectively.

The 1D emission model used is the traveling wave antenna (TWA) emission model proposed in the literature [24, 25]. Following Zheltikov [25], the electric field emitted can be written as follows:

$$E(r,t) = \frac{\mu_0 c}{4\pi r}\frac{\sin\theta}{\xi}\left\{i_2\left[t - \frac{nr}{c}\right]\right.$$
$$\left. - i_2\left[t - \frac{nr}{c} - \frac{L\xi}{u}\right]\right\} \quad 13$$

where $i_2$ is the propagating current, $\xi = 1 - \beta n \cos\theta$, $\beta = u/c$, $L$ is the length of the antenna, $u$ is the speed at which the current propagates, $n$ is the index of refraction at the emission wavelength, and $r$ is the detector distance. A Fourier transform of this equation gives the following expression for the electric field:

$$\hat{E}(\omega,\theta) = \frac{\mu_0 c}{4\pi r} \hat{\iota}_2(\omega) \frac{\sin\theta}{\xi} e^{-i\frac{\omega n}{c}r} \times \left[1 - e^{-i\frac{\omega L}{u}\xi}\right] \quad 14$$

$\hat{\iota}_2(\omega) = \hat{\jmath}_2(\omega)S$ is the frequency profile of the laser-driven current and is a complex variable ($S$ is the cross-sectional area of the filament). Taking the amplitude of the complex variable $\hat{E}(\omega,\theta)$ yields:

$$|\hat{E}(\omega,\theta)| = \frac{\mu_0 c}{2\pi r} |\hat{\iota}_2(\omega)| \sin\left(\frac{\omega L \xi}{2c}\right) \frac{\sin\theta}{\xi} \quad 15$$

This is the expression provided by Zhao *et al* [24]. For the calculations below, $u = c$ and $n = 1$. Given these assumptions, Eqn. 15 becomes:

$$|\hat{E}(\omega,\theta)| = \frac{\mu_0 c}{2\pi r} |\hat{\iota}_2(\omega)| \frac{\sin\theta}{1 - \cos\theta} \times \sin\left(\frac{\omega L}{2c}(1 - \cos\theta)\right) \quad 16$$

In this form, the TWA model is equivalent to the original analytic expression given by Amico *et al* [14] for the energy spectral density emitted per unit solid angle. The TWA model has been previously found to yield reasonable agreement with experimental measurements of the angular distribution of emission – for example, Buccheri and Zhang [26] or Zhao *et al* [24].

Equations 12 and 16 can be used to calculate $\hat{E}(\omega,\theta)$ but require a certain list of inputs : the length of the antenna $L$, cross-sectional area of the antenna $S$, electron density $n_0$, and laser pulse profile $I(\eta)$. These can all be estimated using the beam propagation model, together with Eqn. 5 above which accounts for tunneling ionization. In the results that follow, $L$ is taken to be the FWHM of the axial density distribution on axis ($r = 0$). $S = \pi R^2$ where $R$ is the FWHM of the radial density distribution, measured at the axial location of maximum density. $n_0$ is simply the electron density at the focal point obtained from Eqn. 5. The simple model is fully specified and can be solved.

## IV. RESULTS
### IVA. 3D Axisymmetric THz emission model

For the calculations presented here, the optical pulse has a pulse energy of $\sim 30\ \mu J$ and a pulse length of $50\ fs$ (FWHM of intensity profile), with a carrier wavelength of $800\ nm$. It has an initial diameter of $5.0\ mm$ at the lens based upon the full width at half maximum amplitude of the radial intensity profile and is focused down using a 50-cm focal length lens. A rectangular mesh was used for the current dynamics and emission models. This mesh consisted of $[N_Z, N_R] = [11\,000, 50]$ points, spanning $1.1\ cm$ in the axial dimension and $40\ \mu m$ in the radial dimension. The spatial resolution associated with the mesh is approximately $1.0\ \mu m$ in both the axial and radial dimensions. The same axial grid was retained for the beam propagation model calculation. However, a different radial grid was required to capture all relevant spatial frequencies associated with the focusing dynamics. The radial grid chosen had 5000 points spanning a radius of $1\ cm$ – note that these points were not equally spaced and were chosen based upon the algorithm of Guizar-Sicairos *et al* [20]. This grid provided a radial spatial resolution of roughly $2\ \mu m$. Cubic spline interpolation was used to interpolate the laser field profile and electron density $n_0$ between the two radial grids used for the fluid model and beam propagation models.

FIG 2 shows the evolution in time of the laser field amplitude envelope as it propagates through the domain and the electron density $n_0$ – both along the centerline ($r = 0$) as a function of axial distance. The intensity of the laser pulse increases to $\sim 5 \times 10^{13}\ W/cm^2$ at the focus. An analytic calculation performed using a gaussian beam propagation algorithm is shown for comparison (dotted red line). The difference in intensity between the two calculations is due to Kerr self-focusing[1] and plasma defocusing. These nonlinear effects cause a slight asymmetry in the electron density about the point of maximum intensity. Note that the ionization rate of Mishima *et al* [22] is extremely sensitive to intensity. The increase in intensity due to self-focusing increases the ionization rate by a factor of $\sim 3$. Therefore, an important effect of Kerr self-focusing is to alter the plasma density generated by the propagating pulse. Here, the electron density $n_0$ is seen to increase to $\sim 1.2 \times 10^{17}\ cm^{-3}$. The current at second order $J_2$ accounts for THz emission via the *Transition-Cherenkov* mechanism. FIG 3 shows a snapshot of this current, as well as its irrotational ($J_\parallel$) and solenoidal ($J_\perp$) components obtained by performing a Helmholtz decomposition. The irrotational component has zero curl whereas the solenoidal component has zero divergence. As discussed in the literature, only the solenoidal component ($J_\perp$) radiates [27].

---

[1] A small shift in the location of maximum intensity occurs relative to the analytic Gaussian beam calculation. This shift was on the order of $\sim 250\ \mu m$ or 0.05% of the lens focal length and was removed from the numerical calculation in the figures.

The energy arriving to the detector is obtained by integrating the corresponding intensity over time. FIG 4 shows the angular distribution of the predicted emission for the two cases considered previously. It is seen that the emission is primarily in the forward direction (maximum at 5°), consistent with what is reported in the literature [14]. At 5°, the THz emission lasts approximately $50\ fs$. The energy of $2.6\ fJ/sr$ therefore corresponds to an average power of $\sim 50\ mW/sr$. FIG 4 also shows the frequency content of the emission signal at the maximum angle of emission. It is interesting to note that a prominent peak is observed at low frequency ($\sim 2\ THz$). The source of this peak is a coupling between the non-radiating irrotational current ($J_{\parallel}$) and the radiating solenoidal current ($J_{\perp}$). Recall that the calculation assumes a constant value of $\nu_e$ and only accounts for the irrotational component of $E_2$ in Eqn. 7. With these assumptions, taking the curl of Eqn. 7 results in a term proportional to $\nabla n_0 \times E_2$. This term allows for a coupling of $J_{\parallel}$ – which determines $E_2$ through Poisson's equation – and $J_{\perp}$. This coupling enables some of the frequency content of $J_{\parallel}$ to leak through into $J_{\perp}$ and radiate indirectly. However, it is worth noting that the model does not account for shielding effects. Therefore, we do not have confidence in the predictions of the model at frequencies less than the plasma frequency corresponding to the maximum plasma density (indicated in FIG 4).

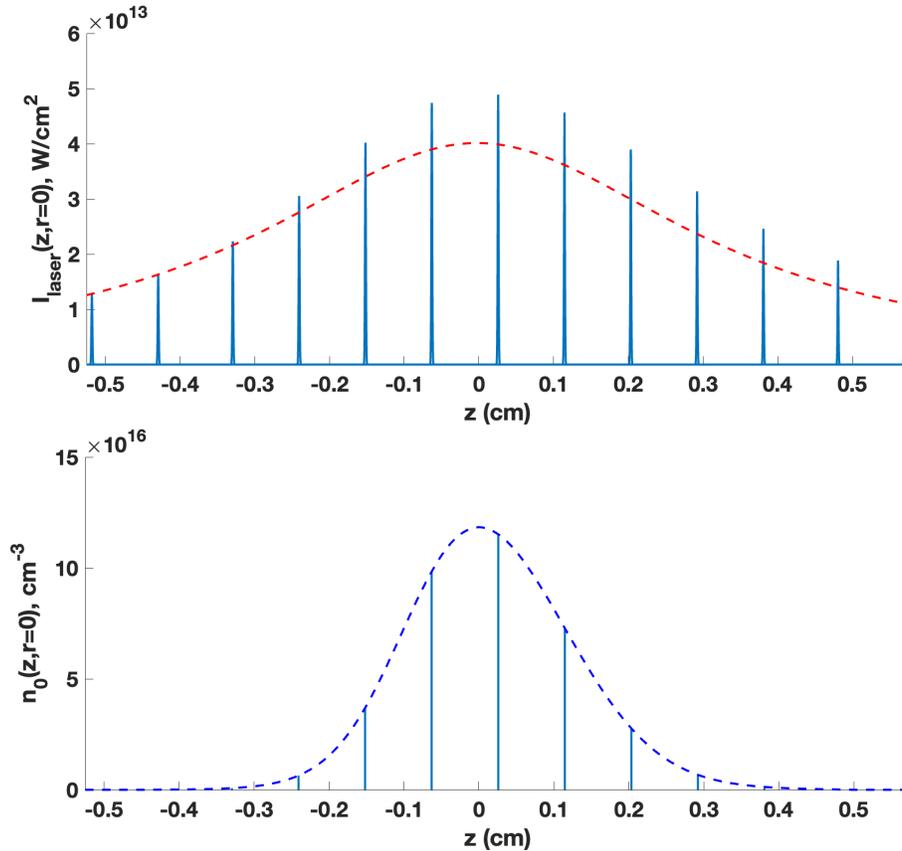

FIG 2: (top) The blue curves correspond to time snapshots of the $50\ fs$ laser amplitude envelope, shown at $3\ ps$ intervals as the pulse moves from left to right. The dashed red curve shows the peak intensity evolution assuming linear propagation of a gaussian beam for comparison. (bottom) The dashed blue line represents the electron density profile $n_0(z, r = 0)$ after the laser pulse has passed through the domain. The solid blue lines indicate the laser pulse location at $3\ ps$ intervals.

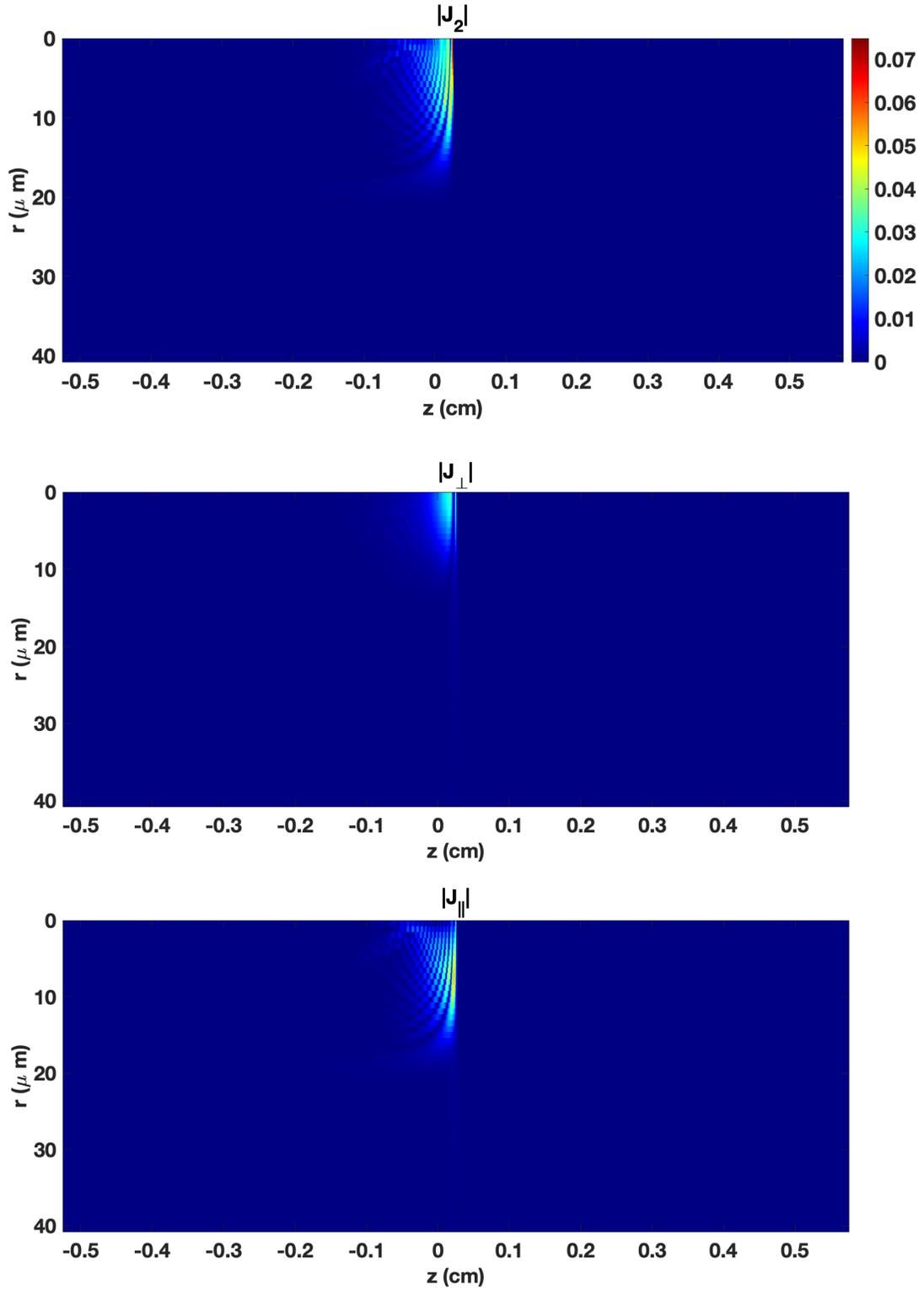

FIG 3: (top) Amplitude of second order current $|J_2|$. The colorbar scale is in units of $mA/\mu m^2$ and applies to all plots. (middle) Amplitude of solenoidal component of second order current. (bottom) Amplitude of irrotational component of second order current.

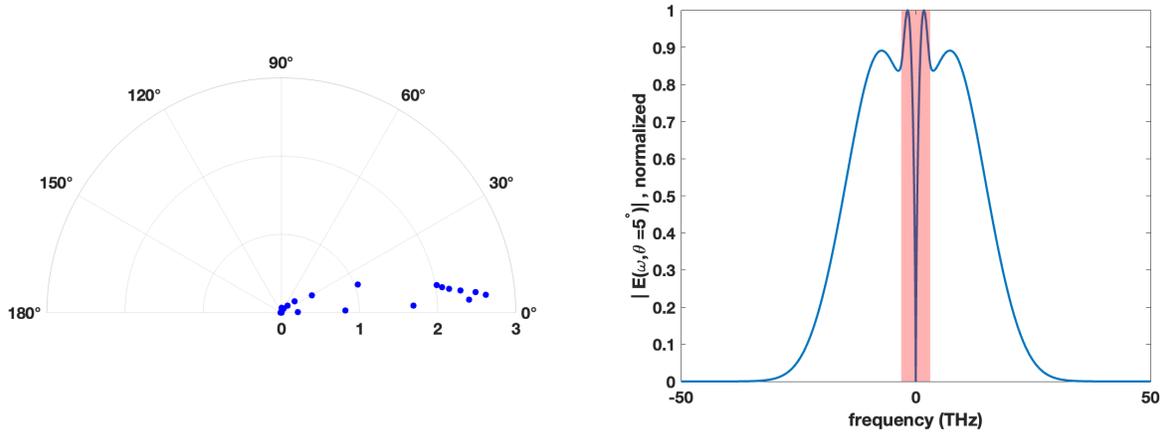

FIG 4: (*left*) Emission diagram for the $f = 50\ cm$ case. The radial axis units on both plots are $fJ/sr$ – to convert to energy per unit area, divide by the square of the detector distance. (*right*) Normalized frequency content of the electric field at the maximum emission angle of 5°. The highlighted zone shows all frequencies that fall within the plasma frequency based upon the maximum electron density.

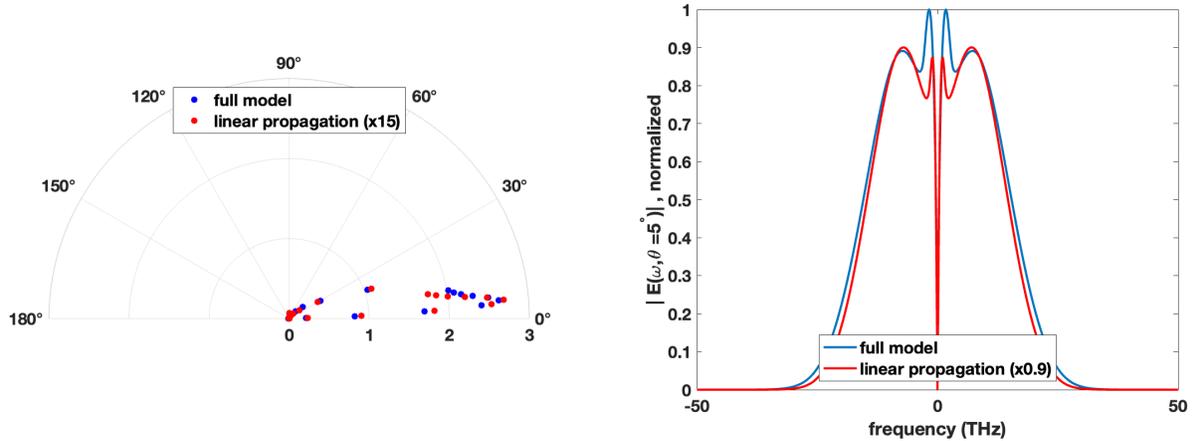

FIG 5: (*left*) Emitted energy as a function of angle ($fJ/sr$), as predicted by both the full model and when assuming linear propagation (no self-focusing or plasma defocusing). (*right*) Normalized frequency content of the electric field at the maximum emission angle of 5°. The linear propagation calculation has been normalized to its maximum and then scaled by a factor of 0.9 – the latter scaling is done to highlight the good agreement at larger frequencies.

One initial goal was to evaluate the effect of nonlinear propagation dynamics on the THz signal. FIG 5 shows a comparison of the 3D axisymmetric model predictions, both with (*full model*) and without (*linear propagation*) accounting for nonlinear propagation effects. For the case considered here, the primary effect is on the amplitude of the THz signal which is strongly impacted by these dynamics. Using a linear propagation model underpredicts the signal amplitude by a factor of ~15. Nonlinear propagation dynamics alter the intensity distribution near the focus, and this alteration leads to strong alterations in the electron density due via the high sensitivity of the tunneling ionization cross-section to intensity. This, in turn, results in an increase in the radiating current amplitudes. However, despite the large sensitivity of the signal amplitude to nonlinear propagation effects, the angular distribution and frequency content of the signal remains comparatively unaffected. At lower frequencies, the linear propagation model fails to capture the structure. However, this discrepancy occurs at or below the plasma frequency and thus shielding will likely be important. This shielding is not accounted for, even in the full model.

An alternate test case was also considered wherein a higher energy pulse was focused down into $N_2$ producing a comparable total electron density (~$1.0 \times 10^{17} cm^{-3}$). The higher energy results in a larger intensity at the focus, and thus the effect of Kerr self-focusing is stronger. For this alternate case, a linear

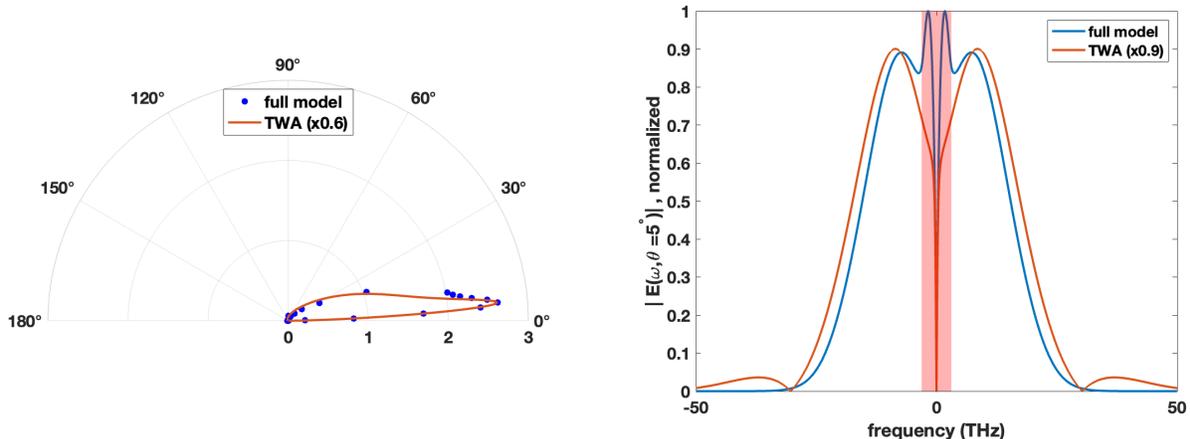

FIG 6: (*left*) Emitted energy as a function of angle ($fJ/sr$), as predicted by both the full model and the simplified TWA. The TWA calculation has been scaled by 0.6 to highlight the good agreement in the angular emission distribution. (*right*) Normalized frequency content of the electric field at the maximum emission angle of 5°. The TWA calculation has been normalized to its maximum and then scaled by a factor of 0.9 – the latter scaling is done to highlight the good agreement at larger frequencies.

Gaussian beam calculation underpredicted the signal amplitude by a factor of ~750. Nonetheless, despite the large difference in amplitude, the frequency content and angular distribution predicted by the linear Gaussian beam calculation were quite close to the predictions of the full calculation.

**IVB. Quasi-1D THz emission model**

As in Section IVA, a grid of $[N_Z, N_R] = [11\,000, 5000]$ points is used for the beam propagation model. This grid spans a domain $0.5\,cm$ long in the axial dimension and $1\,cm$ wide in the radial dimension. This calculation provides the necessary inputs for the calculation detailed in section 0. FIG 6 shows the comparison between the 3D axisymmetric and quasi-1D numerical simulations. The angular emission profile is captured quite well by the traveling wave antenna model – both the emission amplitude and angular distribution. This is consistent with results in the literature, though to our knowledge, only a comparison with the angular distribution has been made. However, it is worth noting that the beam propagation model is necessary to correctly predict the amplitude of the THz emission signal. This is because Kerr self-focusing increases the intensity of the laser pulse at the focus, resulting in a much larger ionization rate. Though the increase in intensity at the focus is only moderate for this case, the ionization cross-section is extremely sensitive to intensity. Meanwhile, the frequency content of the emission signal is well estimated as well. These observations give us confidence in the predictions of the 3D axisymmetric numerical model.

**V. CONCLUSIONS**

Nonlinear propagation effects become more important at higher intensities and/or longer interaction lengths. For the case considered in this paper, the maximum intensity is $\sim 5 \times 10^{13}\,W/cm^2$ and the distance over which nonlinear propagation dynamics have a significant effect is $\sim 0.5\,cm$ (see FIG 2). These nonlinear propagation dynamics have a strong impact on the amplitude of the THz emission signal. However, the impact on the frequency content and angular distribution of the emission is comparatively small. We use a beam propagation model that can be readily coupled to existing models of THz emission in the literature. When used in conjunction with simple 1D models of current dynamics and the Traveling Wave Antenna emission model, the beam propagation model yields estimates of energy, frequency content, and angular distribution of the emission that are in good agreement with more detailed numerical estimations (see FIG 6).

**ACKNOWLEDGMENTS**

This work was partially supported by the Princeton Collaborative Research Facility (PCRF) supported by the U.S. DOE.

**APPENDIX A – PONDEROMOTIVE SOURCE TERM**

The first order current and electric field have magnitudes of $J_1$ and $E_1$, respectively. For linearly polarized light, the corresponding vectors are either parallel or anti-parallel. Without loss of generality, the amplitude of these vectors can be written as follows:

$$J_1 = Re[\bar{J}_1(\mathbf{r}, t) e^{i(k_L z - \omega_L t)}]$$

$$E_1 = Re[\bar{E}_1(r,t)e^{i(k_L z - \omega_L t)}]$$

where $k_L$ and $\omega_L$ are the laser wavenumber and frequency, respectively. Substituting the above relations into Eqn. 6 and simplifying gives:

$$-i\omega_L \bar{J}_1 + \frac{\partial \bar{J}_1}{\partial t} + \nu_e \bar{J}_1 = \frac{e^2 n_0}{m_e} \bar{E}_1$$

The slowly varying envelope approximation is now applied, wherein it is assumed that $|-i\omega_L \bar{J}_1| \gg |\partial \bar{J}_1/\partial t|$. With this approximation, it follows that:

$$\bar{J}_1 = \frac{e^2 n_0}{m_e} \frac{\bar{E}_1}{-i\omega_L + \nu_e}$$

Now consider the first term on the right-side of Eqn. 8 which, as explained by Thiele *et al* [5], represents the effect of the ponderomotive force :

$$l_2^{Pond} = \frac{n_0}{2e} \nabla \left|\frac{J_1}{n_0}\right|^2$$

where

$$|J_1|^2 = \frac{1}{4}[\bar{J}_1 e^{i(k_L z - \omega_L t)} + \bar{J}_1^* e^{-i(k_L z - \omega_L t)}][\bar{J}_1^* e^{-i(k_L z - \omega_L t)} + \bar{J}_1 e^{i(k_L z - \omega_L t)}]$$

and $*$ denotes the complex conjugate. If high frequency terms going as $e^{\pm i 2 k_L z}$ and $e^{\pm i 2 \omega_L t}$ are ignored, the above relation yields:

$$|J_1|^2 \sim \frac{\bar{J}_1 \bar{J}_1^*}{2}$$

and, finally:

$$l_2^{Pond} \sim \frac{n_0}{2e} \nabla \left(\frac{\bar{J}_1 \bar{J}_1^*}{2n_0^2}\right) \qquad 17$$

## APPENDIX B: RADIATION PRESSURE AND IONIZATION SOURCE TERMS

The second term on the right-side of Eqn. 8 accounts for the so-called radiation pressure and ionization sources [5], and is as follows :

$$l_2^{RP} + l_2^{Ion} = \frac{J_1}{e} \times \nabla \times \int_{-\infty}^{t} \frac{J_1}{n_0}\left(\nu_e + \frac{\partial_t n_0}{n_0}\right) dt'$$

Whereas the ponderomotive source term was compatible with an assumption of axisymmetry, only a portion of this source term is compatible. To see this, it is easiest to use a Cartesian coordinate system. Furthermore, the scalar function $f(x,y,z)$ will be defined such that:

$$f(x,y,z) = \frac{1}{en_0}\left(\nu_e + \frac{\partial_t n_0}{n_0}\right)$$

and

$$l_2^{RP} + l_2^{Ion} = J_1 \times \nabla \times \int_{-\infty}^{t} f J_1 \, dt'$$

Recall that $E_1$ is the laser electric field and, without loss of generality for linearly polarized laser pulses, let $E_1 = E_1 \hat{x}$. Inspection of Eqn. 6 shows that $J_1 = J_1 \hat{x}$. It follows that:

$$l_2^{RP} + l_2^{Ion} = \left[J_1 \frac{\partial}{\partial z}\left(\int_{-\infty}^{t} f J_1 \, dt'\right)\right]\hat{z} + \left[J_1 \frac{\partial}{\partial y}\left(\int_{-\infty}^{t} f J_1 \, dt'\right)\right]\hat{y}$$

The second term on the right-side of the above equation is not compatible with an assumption of axisymmetry and will therefore be ignored. Letting $l_2 = (l_2^{RP} + l_2^{Ion})\hat{z}$, the remaining term can be expanded as follows:

$$l_2^{RP} + l_2^{Ion} = J_1 \int_{-\infty}^{t} \left[\frac{\partial f}{\partial z}J_1 + f\frac{\partial J_1}{\partial z}\right] dt'$$

As in Appendix A, $J_1 = Re[\bar{J}_1 e^{i(k_L z - \omega_L t)}]$. Applying the slowly varying envelope approximation yields:

$$\frac{\partial J_1}{\partial z} \sim \frac{1}{2}[ik_L \bar{J}_1 e^{i(k_L z - \omega_L t)} + c.c.]$$

where $c.c.$ denotes '*complex conjugate*'. It follows that:

$$l_2^{RP} + l_2^{Ion} \sim \frac{1}{2}[\bar{J}_1 e^{i(k_L z - \omega_L t)} + c.c.] \times \int_{-\infty}^{t}\left[\left(\frac{1}{2}\frac{\partial f}{\partial z}\bar{J}_1 e^{i(k_L z - \omega_L t')} + c.c.\right) + \left(\frac{ik_L f}{2}\bar{J}_1 e^{i(k_L z - \omega_L t')} + c.c.\right)\right] dt'$$

Two new functions $A$ and $B$ are now defined as follows:

$$A = \frac{1}{2}\int_{-\infty}^{t} \frac{\partial f}{\partial z}\bar{J}_1 e^{-i\omega_L t'} \, dt'$$

$$B = \int_{-\infty}^{t} \frac{ik_L f}{2}\bar{J}_1 e^{-i\omega_L t'} \, dt'$$

so that:

$$l_2^{RP} + l_2^{Ion} \sim \frac{1}{2}[\bar{J}_1 e^{i(k_L z - \omega_L t)} + c.c.][(Ae^{ik_L z} + c.c.) + (Be^{ik_L z} + c.c.)]$$

Expanding out the product and neglecting high frequency terms that go as $e^{\pm i 2 k_c z}$ gives:

$$l_2^{RP} + l_2^{Ion} \sim \left(\frac{1}{2}\bar{J}_1 A^* e^{-i\omega_L t} + c.c.\right) + \left(\frac{1}{2}\bar{J}_1 B^* e^{-i\omega_L t} + c.c.\right)$$

Meanwhile, terms $A$ and $B$ can be simplified by applying the slowly varying envelope approximation. This will be illustrated using term $A$. First, let $\Gamma(t) = (1/2)\bar{J}_1(\partial f/\partial z)$ and $\hat{\Gamma}(\omega)$ be its Fourier transform so that

$$\Gamma(t) = \int_{-\infty}^{\infty} \hat{\Gamma}(\omega)\, e^{i\omega t} d\omega$$

and

$$\Gamma'(t) = \int_{-\infty}^{\infty} i\omega\, \hat{\Gamma}(\omega)\, e^{i\omega t} d\omega$$

One way to evaluate term $A$ is by substituting directly the Fourier transform of $\Gamma(t)$ into the expression as follows:

$$A = \int_{-\infty}^{t} \left[ \int_{-\infty}^{\infty} \hat{\Gamma}(\omega) e^{i(\omega - \omega_L) t'} d\omega \right] dt' \qquad 18$$

Another way of evaluating the integral is using integration by parts, which gives:

$$A = -\Gamma(t) \frac{e^{-i\omega_L t}}{i\omega_L} + \frac{1}{i\omega_L} \int_{-\infty}^{t} \Gamma'(t') e^{-i\omega_L t'} dt'$$

Substituting the above expression of $\Gamma'(t)$ into the integral and rearranging gives:

$$A = -\Gamma(t) \frac{e^{-i\omega_L t}}{i\omega_L} + \int_{-\infty}^{t} \left[ \int_{\infty}^{\infty} \frac{\omega}{\omega_L} \hat{\Gamma}(\omega) e^{i(\omega - \omega_L) t'} d\omega \right] dt' \qquad 19$$

Visual inspection of Eqns. 18 and 19, and noting that $\hat{\Gamma}(\omega)$ is non-zero only for $|\omega/\omega_L| \ll 1$ (slowly varying envelope approximation), shows that:

$$A \sim -\Gamma(t) \frac{e^{-i\omega_L t}}{i\omega_L} = -\frac{1}{2} \frac{\partial f}{\partial z} \bar{J}_1 \frac{e^{-i\omega_L t}}{i\omega_L}$$

$$B \sim -\frac{i k_L f}{2} \bar{J}_1 \frac{e^{-i\omega_L t}}{i\omega_L}$$

for both $A$ and $B$. These can be substituted into the above relation for $l_2^{RP} + l_2^{Ion}$ to obtain a simplified expression.

## APPENDIX C: DERIVATION OF 1D EXPRESSION FOR $l_2$

Of the four terms on the right-hand side of Eqn. 8, only three were accounted for in the 3D axisymmetric model – the third term was ignored because it was not consistent with the assumption of axisymmetry. Meanwhile, the last term on the right-hand side is not expected to contribute significantly to the radiation as this term is irrotational (zero curl). There would need to be a coupling between the irrotational and solenoidal current components for this term to have any impact on the radiation. With these observations, only the first two terms are retained for this simplified 1D model. These terms account for the ponderomotive and radiation pressure forces that drive the current, as well as the ionization that also contributes to the current. The 1D expression for the ponderomotive source term follows directly from Eqn. 17:

$$\boldsymbol{l}_2^{Pond} \sim \frac{n_0}{2e} \frac{d}{dz} \left( \frac{\bar{J}_1 \bar{J}_1^*}{2 n_0^2} \right) \hat{\boldsymbol{z}}$$

Thiele et al [5] provide simplified expressions for the radiation pressure ($\boldsymbol{l}_2^{RP}$) and ionization ($\boldsymbol{l}_2^{Ion}$) source terms within the context of a 1D model. These are as follows: $\boldsymbol{l}_2^{RP} = -(n_0 \nu_e / e\, c) |J_1/n_0|^2\, \hat{\boldsymbol{z}}$ and $\boldsymbol{l}_2^{Ion} = -(\partial_\eta n_0 / e\, c) |J_1/n_0|^2\, \hat{\boldsymbol{z}}$, where $\eta = t - z/c$. Applying the slowly varying approximation to these expressions gives:

$$\boldsymbol{l}_2^{RP} = -\frac{n_0 \nu_e}{ec} \frac{\bar{J}_1 \bar{J}_1^*}{2 n_0^2} \hat{\boldsymbol{z}}$$

$$\boldsymbol{l}_2^{Ion} = -\frac{\partial_\eta n_0}{ec} \frac{\bar{J}_1 \bar{J}_1^*}{2 n_0^2} \hat{\boldsymbol{z}}$$

Using the expressions in Appendix A to replace $J_1$ with $E_1$ in the above expressions, noting that $I = (1/2) \epsilon_0 c \bar{E}_1 \bar{E}_1^*$, and assuming that $I(z,t) = I(\eta = t - z/c)$ gives:

$$\boldsymbol{l}_2(\eta) = \boldsymbol{l}_2^{Pond} + \boldsymbol{l}_2^{RP} + \boldsymbol{l}_2^{Ion} = l_2(\eta)\, \hat{\boldsymbol{z}}$$

$$= -\frac{1}{e \epsilon_0 c^2} \left( \frac{e^2}{m_e} \right)^2 \frac{1}{\omega_L^2 + \nu_e^2} \left[ \frac{n_0}{2} \frac{dI}{d\eta} + \nu_e n_0 I + I \frac{dn_0}{d\eta} \right] \hat{\boldsymbol{z}}$$

The above expression also assumes that $\nu_e$ is constant.